\documentclass{ws-ijmpe}

\newcommand{\vm}[1]{\overrightarrow{#1}}
\newcommand{\vmp}[1]{\overrightarrow{#1}^{\prime}}
\newcommand{\nur}{\hat{n}_{\uparrow}(\vm{r})}
\newcommand{\ndr}{\hat{n}_{\downarrow}(\vm{r})}
\newcommand{\nurp}{\hat{n}_{\uparrow}(\vmp{r})}
\newcommand{\ndrp}{\hat{n}_{\downarrow}(\vmp{r})}
\newcommand{\nsum}{n_{\uparrow}(\vm{r})+n_{\downarrow}(\vm{r})+n_{\uparrow}(\vmp{r})+n_{\downarrow}(\vmp{r})}

\begin{document}

\markboth{G. Wlaz{\l}owski, P. Magierski}{Quantum Monte-Carlo study of strongly correlated dilute fermionic gases}

\catchline{}{}{}{}{}

\title{QUANTUM MONTE CARLO METHOD APPLIED TO STRONGLY CORRELATED DILUTE FERMI GASES WITH FINITE EFFECTIVE RANGE.}

\author{\footnotesize GABRIEL WLAZ{\L}OWSKI and PIOTR MAGIERSKI}

\address{Faculty of Physics, Warsaw University of Technology, ul. Koszykowa 75\\
00-662 Warsaw,
Poland\\
e-mail: gabrielw@if.pw.edu.pl, magiersk@if.pw.edu.pl}

\maketitle

\begin{history}
\received{(received date)}
\revised{(revised date)}
\end{history}

\begin{abstract}
We discuss the Auxiliary Field Quantum Monte Carlo (AFQMC) method applied to dilute neutron matter
at finite temperatures. We formulate the discrete Hubbard-Stratonovich transformation for
the interaction with finite effective range which is free from the sign problem. The AFQMC results
are compared with those obtained from exact diagonalization
for a toy model. Preliminary calculations of energy and 
chemical potential as a function of temperature are presented.
\end{abstract}

\section{Introduction}
Properties of dilute and degenerate Fermi gases with large scattering lengths are of particular interest
in the context of cold fermionic atoms and dilute neutron matter. 
In the case of trapped fermionic atoms, due to very low density, the scattering length is the 
only parameter which determines an atom-atom interaction. 
Taking advantage of Feshbach resonances it can be tuned, using  external magnetic field, to the
values which greatly exceed the average interparticle distance. This limit which is dubbed the unitary
regime exhibits universal properties and has been the subject of intensive theoretical effort in the last
couple of years (see \cite{gps} and references therein). Similarly, neutron matter
at densities corresponding to $ k_F \lesssim 0.6$ fm$^{-1}$ is a dilute
system of fermions in a sense that the average distance between particles is much larger than
the range of the neutron-neutron interaction. It is superfluid and 
its physics is determined by two parameters: scattering
length $a_{S}$ and effective range $r_{\mbox{eff}}$ in $^{1}S_{0}$ neutron-neutron channel. 
The influence of other channels
as well as of three-body forces is marginal and can be neglected in this density range\cite{bm}.
The properties of dilute neutron matter are crucial for understanding the structure and thermal evolution
of neutron star crust\cite{pr}. They can also provide constraints for parameters of nuclear energy
density functional.

Despite of these similarities the important qualitative difference between 
dilute neutron matter and cold atomic gases arises from the fact that in the former case 
the nonzero effective range cannot be neglected. Therefore contrary to cold atomic gases, physics of dilute neutron matter cannot be captured by simple contact, delta-like interaction (except for very low densities $k_F \lesssim 0.06 $fm$^{-1}$).

Since Fermi gas with long scattering length is an example of strongly correlated system, non-perturbative
methods based on Monte Carlo approaches are often used to determine its properties.
There exist various calculations at zero temperature \cite{1,2,3,4,5,6,7,8,9,10,11}.
Here we present selected aspects of AFQMC approach at finite temperatures,
which is able to generate fully non-perturbative results for dilute neutron matter without sign problem.

\section{Theoretical approach}

Neutron-neutron interaction in dilute neutron matter can be described by small number of parameters characterizing the low energy physics of two-body collisions. In this limit the main contribution to the scattering amplitude comes from s-wave scattering and is determined by two parameters: scattering length $a_{S}$ and effective range $r_{\mbox{eff}}$ i.e.:
\begin{equation}
 f(p)=\dfrac{1}{-1/a_{S} + \frac{1}{2}r_{\mbox{eff}}p^{2}-ip},
\end{equation} 
where $a_{S}\approx-18.5fm$ and $r_{\mbox{eff}}\approx2.8fm$.

To capture low energy physics of dilute neutron matter we consider the Hamiltonian of the form:
\begin{eqnarray}
  \hat{H}&=&\sum_{\lambda=\pm}\int  d^{3}\vm{r} \hat{\psi}^{\dagger}_{\lambda}(\vm{r})\left ( \dfrac{-\nabla^{2}}{2m} \right ) \hat{\psi}_{\lambda}(\vm{r}) \nonumber\\
  &&+\dfrac{1}{2}\sum_{\lambda,\lambda^{\prime}=\pm}\int  d^{3}\vm{r}d^{3}\vmp{r} \,V(\vm{r}-\vmp{r}) \hat{\psi}^{\dagger}_{\lambda}(\vm{r}) \hat{\psi}^{\dagger}_{\lambda^{\prime}}(\vmp{r}) \hat{\psi}_{\lambda^{\prime}}(\vmp{r}) \hat{\psi}_{\lambda}(\vm{r}) \label{inter_r}, 
\end{eqnarray} 
where the field operators obey the fermionic anticommutation relations $ \{ \hat{\psi}^{\dagger}_{\lambda}(\vm{r}),\hat{\psi}_{\lambda^{\prime}}(\vmp{r})\} = \delta_{\lambda\lambda^{\prime}}\delta(\vm{r}-\vmp{r})$ and $\lambda$ denotes the spin degree of freedom. 
Throughout this work we shall use units in which Planck's and Boltzmann's constants are equal
to one.
The interaction has the form:
\begin{equation}
 V(\vm{r}-\vmp{r})=\left\lbrace \begin{array}{ll}
                      6g,&\vm{r}-\vmp{r}=0\\
                      g,&\vm{r}-\vmp{r}\in \mathcal{N}_{b}\\
                      0, &otherwise.
                     \end{array}\right.  , 
\end{equation}
where $\mathcal{N}_{b}=\{(b,0,0),(-b,0,0),(0,b,0),(0,-b,0),(0,0,b),(0,0,-b) \}$.\\
This choice has been motivated, as will become clear later, by the special form of discrete Hubbard-Stratonovich (H-S) transformation. It imposes limitations on the density range
of neutron matter which can be described by this particular interaction.
Clearly the interaction depends on two parameters: $g$ and $b$. Solving the
equation for $T$-matrix elements:
\begin{equation}
 T_{\vm{p}\vmp{p}}=V_{\vm{p}\vmp{p}}+\int \dfrac{d^{3}\vm{k}}{(2\pi)^{3}}V_{\vm{p}\vm{k}}G_{\vmp{p}\vm{k}}T_{\vm{k}\vmp{p}}\label{eqn:Tmatrix},
\end{equation}
where $G$ is the free particle propagator and using the relation:
\begin{equation}
 -\dfrac{4\pi}{m}T^{-1}_{\vm{p}\vm{p}}\approx-\dfrac{1}{a_{S}}+\dfrac{1}{2}r_{\mbox{eff}}p^{2}-ip+O(p^{3})
\end{equation}
we can adjust parameters $g$ and $b$ to reproduce scattering length $a_{S}$ and effective range $r_{\mbox{eff}}$ of the neutron-neutron interaction. 
Still however the regularization procedure is required, which can be
determined by introducing the momentum cut-off $p_{cut}$.
This prescription sets to zero all two-body matrix elements, if the relative momentum of two particles exceeds a given  momentum cut-off.

In order to compute thermal properties of dilute neutron matter we place the system on a 3D spatial lattice.  The lattice spacing $b$ and size $L = N_s b$ introduce natural ultraviolet (UV) and 
infrared (IR) momentum cut-offs given by $p_{cut}=\pi/b$ and $p_0=2\pi/L$, respectively. 
The momentum space has the shape of a cubic lattice, 
with size $2\pi/b$ and spacing $2\pi/L$.
To simplify the analysis, however, we place a spherically symmetric UV cut-off, including  
only momenta satisfying $p \le p_{cut} $.

To evaluate expectation values of observables we have used a path integral representation of partition function:
\begin{eqnarray}
 Z(\beta,\mu) &=& {\mathrm{Tr}} \left \{ \prod_{k=1}^{N_\tau} \exp [-\tau (\hat{H}-\mu \hat{N})] \right \},\\
 O(\beta,\mu) &=& 
\dfrac{1}{Z(\beta,\mu)}{\mathrm{Tr}} \; \left \{ \hat{O}\prod_{k=1}^{N_\tau} \exp [-\tau (\hat{H}-\mu \hat{N})] \right \},
\end{eqnarray} 
where $\beta = 1/T = N_\tau \tau$, $T$ is the temperature and $\mu$ - the chemical potential, $\hat{N}$ - the particle number operator and $\hat{O}$ is a quantity of interest. The propagator $e^{-\tau (\hat{H}-\mu \hat{N})}$ is decomposed using 
the second order expansion\cite{bdm1,bdm2}:
\begin{equation}\label{eq:fact-Exp}
\exp[ -\tau(\hat{H}-\mu \hat{N})] \approx
\exp \left [ -\frac{\tau (\hat{T}-\mu \hat{N})}{2}\right ]
\exp(-\tau \hat{V} )
\exp \left [ -\frac{\tau (\hat{T}-\mu \hat{N})}{2}\right ],
\end{equation}
where $\hat{T}$ is the kinetic energy operator. To express the many body operator $e^{-\tau \hat{V}}$ by a sum of one body operators we have extended the discrete H-S transformation introduced by Hirsch\cite{hirsch}. Namely, one may rewrite the interaction in the form:
\begin{eqnarray}
  \hat{V}&=&\dfrac{g}{2}\sum_{\vm{r}-\vmp{r}\in\mathcal{N}_{b}}(\nur\nurp+\nur\ndrp+\ndr\nurp+\nonumber\\
             &&+\ndr\ndrp+\nur\ndr+\nurp\ndrp), \label{inter_decomp}
\end{eqnarray}
where $\hat{n}_{\lambda}(\vm{r})=\hat{\psi}^{\dagger}_{\lambda}(\vm{r})\hat{\psi}_{\lambda}(\vm{r})$. 
Note that in the coordinate representation each element of the above sum
can take the following values:
\begin{equation}
 \left\lbrace \begin{array}{ll}
                 0,& \textrm{iff } \nsum=0\\
                 0,& \textrm{iff } \nsum=1\\
                 g/2,& \textrm{iff } \nsum=2\\
                 3g/2,& \textrm{iff } \nsum=3\\
                 6g/2,& \textrm{iff } \nsum=4
                \end{array}\right. 
\end{equation}
where $n_{\lambda}(\vm{r})$ denotes eigenvalue of density operator. 
This allows us to introduce the H-S transformation of the form:
\begin{equation}\label{eq:hs-transformation} 
 e^{-\tau\widehat{V}}=\prod_{\vm{r}-\vmp{r}\in\mathcal{N}_{b}}\dfrac{1}{k}\sum_{i=1}^{k}e^{\sigma_{i}(\vm{r},\vmp{r})[\nur+\ndr+\nurp+\ndrp]}
\end{equation}
where values of $\sigma_{i}$ and $k$ have to fulfil the equations:
\begin{equation}
\left\lbrace \begin{array}{l}
               \frac{1}{k}\sum_{i=1}^{k}e^{\sigma_{i}}=1\\
               \frac{1}{k}\sum_{i=1}^{k}e^{2\sigma_{i}}=e^{-\frac{\tau g}{2}}\\
               \frac{1}{k}\sum_{i=1}^{k}e^{3\sigma_{i}}=e^{-\frac{3\tau g}{2}}\\
               \frac{1}{k}\sum_{i=1}^{k}e^{4\sigma_{i}}=e^{-\frac{6\tau g}{2}}.
              \end{array} \right. 
\end{equation} 
This set of equations has a solution if $k\geqslant 7$ and $g<0$ (attraction).

The partition function can be expressed in the form:
\begin{eqnarray}
Z(\beta,\mu) &=& 
\int \prod_{\tau_{j}}\prod_{\vm{r}-\vmp{r}\in\mathcal{N}_{b}}{\cal{D}}\sigma(\vm{r},\vmp{r},\tau_{j})\,
{\mathrm{Tr}}\ {\cal{\hat{U}}}(\{ \sigma \}),\nonumber \\
{\cal{\hat{U}}}(\{ \sigma \})&=&{\mathrm{T}}_\tau\exp  \{ -\tau [\hat{h}(\{\sigma\})-\mu ]\},
\end{eqnarray}
where $\mathrm{T}_{\tau}$ defines the time-ordered product and $\hat{h}(\{\sigma\})$ denotes the
one-body Hamiltonian. Consequently the expectation value of an arbitrary operator takes the form\cite{bdm1,bdm2}:
\begin{equation} \label{eq:ham-T}
O(\beta,\mu) = \langle\hat{O}\rangle=
\int \frac{\prod{\cal{D}}\sigma(\vm{r},\vmp{r},\tau_{j})\,
{\mathrm{Tr}}\;{\cal{\hat{U}}}(\{ \sigma \})}{Z(\beta,\mu)}
\; \frac{ {\mathrm{Tr}}\; \hat{O}{\cal{\hat{U}}}(\{ \sigma \})}
{{\mathrm{Tr}}\; {\cal{\hat{U}}}(\{ \sigma \})}.
\end{equation}
The many-fermion problem has thus been reduced to the typical AFQMC problem. 
Consequently the standard Metropolis algorithm can be applied, using ${{\mathrm{Tr}}\; {\cal{\hat{U}}}(\{ \sigma \})}$ as a measure\cite{kdl}.

To prove that the measure is positive it is sufficient to note that the operator ${\cal{\hat{U}}}(\{ \sigma \})$
is invariant with respect to time reversal. This is due to the special form of expression (\ref{eq:hs-transformation})
and the fact that the sigma field is real.
Therefore the single particle spectrum of ${\cal{\hat{U}}}(\{ \sigma \})$
consists of pairs which are conjugate to each other. Clearly 
${{\mathrm{Tr}}\; {\cal{\hat{U}}}(\{ \sigma \})} = \det(1+{\cal{\hat{U}}}(\{ \sigma \})) = \prod_{i} |(1+u_{i})|^{2}\geqslant0$.
In particular if the single particle basis is chosen in such a way that the time reversed partners
correspond to spin-up and spin-down particles then the matrix representation of $\cal{\hat{U}}$ takes the form:
\begin{equation}
 \cal{U}=\left( \begin{array}{cc}
                              \cal{U}_{\uparrow\uparrow} & 0\\
                              0                          & \cal{U}_{\downarrow\downarrow}
                              \end{array}\right)
\end{equation} 
where ${\cal{U}_{\uparrow\uparrow}}= {\cal{U}_{\downarrow\downarrow}}^{*}$. 
Matrix elements between states with different spin projections are equal to zero. Finally one gets
\begin{equation}
  {\mathrm{Tr}}\; {\cal{\hat{U}}}=\det [1 + {\cal{U}}]=\det [1 + {\cal{U}_{\uparrow\uparrow}}]\det [1 + {\cal{U}_{\uparrow\uparrow}^{*}}]=|\det [1 + {\cal{U}_{\uparrow\uparrow}}]|^{2}\geqslant 0.
\end{equation}
Hence the lack of the sign
problem is a direct consequence of the time-reversal symmetry which is preserved by
each term in the sum (\ref{eq:hs-transformation}).

\section{Numerical results}

In order to verify the correctness of our code we have performed several tests. As a first check the thermodynamics of a free gas was reproduced for $g = 0$.

To check our results for an interacting system we have diagonalized 
the Hamiltonian (2) exactly, restricting the 
single-particle Hilbert space 
to the lowest 7 states. It corresponds to the Fock space of
$2^{14}$ states. The test provided an estimate for the size of the imaginary time step. 
Comparison of AFQMC results with the exact ones for selected observables at fixed temperature 
is given in the table \ref{tab:comparion}.

\begin{table}[pt] 
\tbl{Comparison of AFQMC results with exact diagonalization in a restricted space (see text for details). $T/\epsilon_{F}=0.58$, $\mu/\epsilon_{F}=0.41$.\label{tab:comparion}}
{\begin{tabular}{@{}ccc@{}} \toprule
Observable & AFQMC result & Exact result \\ 
& (1000 samples) & \\ \colrule
$\langle\hat{T}\rangle$ & 0.32 & 0.31\\ 
$\langle\hat{V}\rangle$ & -0.088 & -0.086\\
$\langle\hat{H}\rangle$ & 0.23 & 0.23 \\
$\langle\hat{N}\rangle$ & 5.58 & 5.49 \\ \botrule

\end{tabular}}
\end{table}

To perform calculations for a neutron matter we have used the box with
$N_{s}=8$ and lattice constant $b=3.21$fm. The neutron mass was fixed at $939$ MeV. 
The chemical potential was chosen in such a way to keep the total number of particles equal to 55. This corresponds to the density of neutron matter $\rho\cong 0.02\rho_{0}$ where $\rho_{0}$ is the saturation density of nuclear matter. For this density we have varied temperatures from $0.06\epsilon_{F}$ ($0.26$ MeV) to $1.0\epsilon_{F}$ ($3.8$ MeV) where $\epsilon_{F}$ is the Fermi energy. For the lowest temperature we have used $2360$ imaginary time steps while for the highest temperature only $216$. In all runs the single-particle occupation probabilities for the highest energy states were below $0.01$ at all temperatures. At low temperatures the Singular Value Decomposition technique was used to avoid instabilities of the algorithm.

\begin{figure}[th]
\centerline{\psfig{file=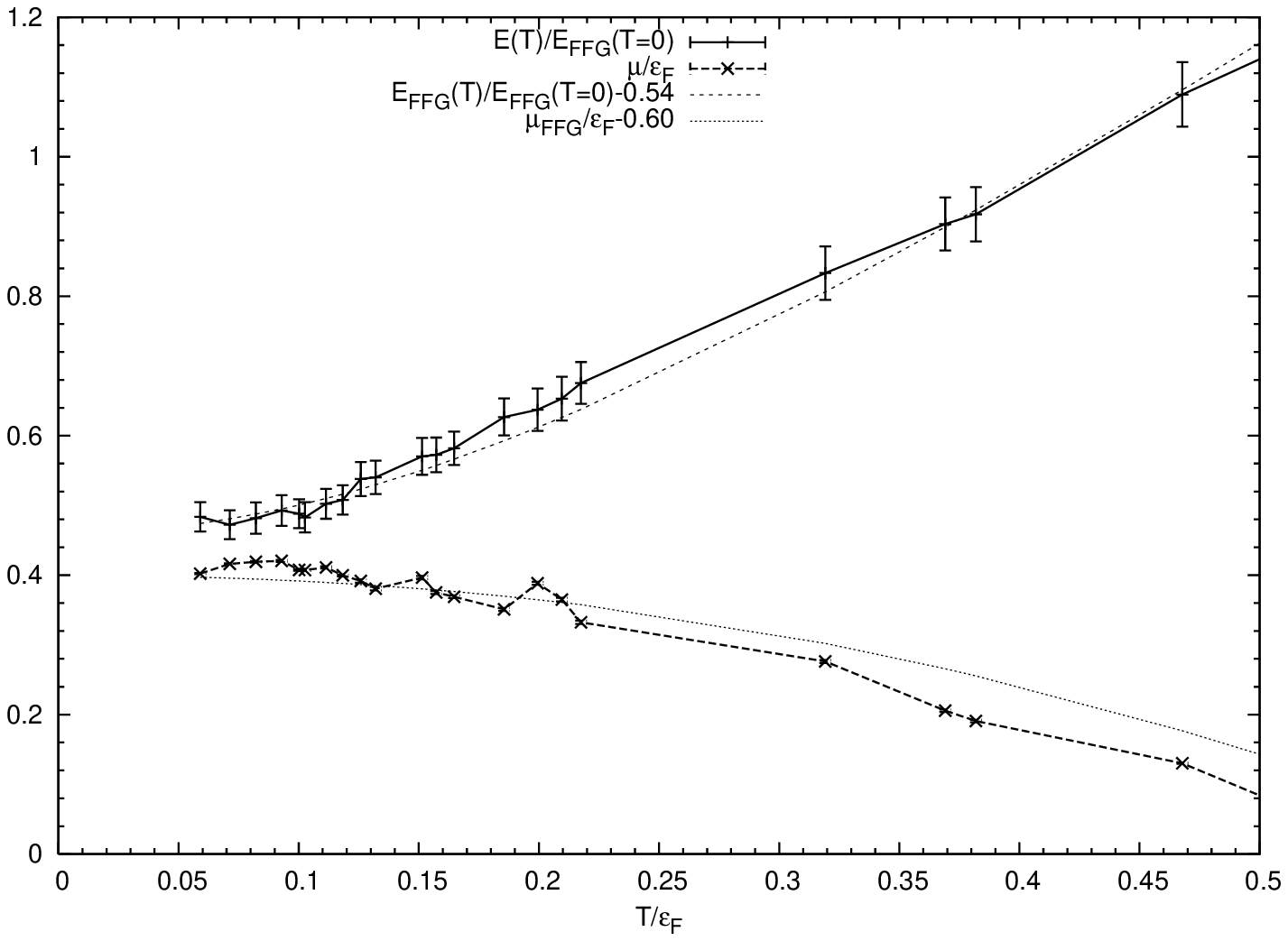,width=7.3cm,angle=-90}}
\vspace*{8pt}
\caption{Energy and chemical potential as a function of temperature for neutron matter at density $\rho\cong 0.02\rho_{0}$. Dotted line denotes energy and chemical potential for free Fermi 
(Free Fermi gas results were shifted by constant values- see text for details.).}\label{fig:eos} 
\end{figure}

Figure \ref{fig:eos} presents an example of the results for the energy and chemical potential as a function of temperature.  
The energy and chemical potential versus temperature for the free Fermi gas at the same particle density 
has also been plotted (dotted line). The free Fermi gas results were shifted by constant values $-0.54$ and $-0.60$ for 
the energy and the chemical potential, respectively.

For our lowest temperature $0.06\epsilon_{F}$ ($0.26$ MeV) we obtained the energy (relative to
the energy of the free Fermi gas $E_{FFG}$)
$E/E_{FFG}=0.48\pm0.02$ ($E/N=1.28$ MeV). This value is in good agreement with recent 
fixed-node quantum Monte Carlo calculations\cite{10}. The chemical potential at this temperature was
calculated to be $\mu/\epsilon_{F}=0.40$ ($1.77$ MeV).

\section{Conclusions}
In this work we have introduced the simple effective interaction which is capable to describe properties of dilute neutron matter. For this interaction we have constructed Hubbard-Stratonovich transformation which yield to positive measure and allows to perform Monte Carlo calculations without fermionic sign problem.
We have performed the fully non-perturbative calculations for dilute neutron matter of density $\rho\cong 0.02\rho_{0}$ at finite temperatures.

\section*{Acknowledgments}
We thank Aurel Bulgac for discussions concerning discrete H-S transformation.
Support from the Polish Ministry of Science under contract No. N N202 328234
is gratefully acknowledged. Use of computers at the Interdisciplinary Centre for Mathematical and Computational Modelling (ICM) at Warsaw University is also gratefully acknowledged.

\end{document}